\documentclass{article}

\usepackage{arxiv}

\usepackage[utf8]{inputenc} 
\usepackage[T1]{fontenc}    
\usepackage{hyperref}       
\usepackage{url}            
\usepackage{booktabs}       
\usepackage{amsfonts}       
\usepackage{nicefrac}       
\usepackage{microtype}      
\usepackage{lipsum}

\usepackage{bm}
\usepackage{amsmath}
\usepackage{amssymb}
\usepackage{amsfonts}
\usepackage{graphicx}
\usepackage{algorithm}
\usepackage{algorithmic}
\usepackage{comment}
\def\vec#1{\mathbf{#1}}

\title{Data-driven End-to-end Learning of Pole Placement Control for Nonlinear Dynamics via Koopman Invariant Subspaces}

\author{
  Tomoharu Iwata\\
  NTT Communication Science Laboratories, Japan\\
  \AND
  Yoshinobu Kawahara\\
  Institute of Mathematics for Industry, Kyushu University, Japan\\
  Center for Advanced Intelligence Project, RIKEN, Japan\\
}
\date{}

\begin{document}
\maketitle

\begin{abstract}
We propose a data-driven method for controlling the frequency and convergence rate of black-box nonlinear dynamical systems based on the Koopman operator theory. With the proposed method, a policy network is trained such that the eigenvalues of a Koopman operator of controlled dynamics are close to the target eigenvalues. The policy network consists of a neural network to find a Koopman invariant subspace, and a pole placement module to adjust the eigenvalues of the Koopman operator. Since the policy network is differentiable, we can train it in an end-to-end fashion using reinforcement learning. We demonstrate that the proposed method achieves better performance than model-free reinforcement learning and model-based control with system identification.
\end{abstract}

\section{Introduction}

Controlling dynamics is important in various fields.
Pole placement, or full state feedback, 
is a fundamental control method~\cite{brasch1970pole},
where a system is controlled such that it has desired eigenvalues.
The eigenvalues represent the frequency and convergence rate of the system.
Although pole placement has been successfully used
for controlling linear dynamical
systems~\cite{chilali1999robust,bevly2007cascaded}, it is inapplicable to nonlinear systems directly.

Recently,
the Koopman operator theory~\cite{koopman1931hamiltonian,mezic2005spectral}  
has been used for data-driven analysis and control of nonlinear systems in a wide variety of applications.
In this theory, a nonlinear dynamical system is lifted to the corresponding linear one
in a possibly infinite-dimensional space by embedding states using a nonlinear function.
By finding such lifted space, which is called a Koopoman invariant subspace~\cite{takeishi2017learning},
we can analyze and control nonlinear systems using various methods developed
for linear systems~\cite{morton2018deep,li2019learning}.
For example, the frequency and convergence rate of a nonlinear system
can be identified with the eigenvalues of a Koopman operator in the lifted space.
Most of the existing control methods based on the Koopman operator theory consist of
two
steps~\cite{han2020cdc,li2019learning,korda2018linear,brunton2016koopman}.
In the first step, a dynamical system is identified, which includes
the estimation of the embedding function and the linear dynamics in the Koopman invariant
subspace~\cite{kawahara2016dynamic,takeishi2017learning,lusch2018deep,iwata2020neural}.
In the second step, a controller is optimized in the Koopman invariant space using the identified system.
Since the two steps are separated, the error accumulated in the system identification cannot be corrected in the second step.
The identified system is not necessarily optimal when controlled.

To alleviate such problems,
end-to-end learning of controllers based on reinforcement learning~\cite{mnih2015human,ohnishi2021koopman}
has been used, where controllers are directly optimized without separated system identification.
Although many model-free reinforcement learning methods have been proposed,
they typically require many training data since they do not model the dynamics.
To improve sample efficiency, a number of methods that combine end-to-end learning and model-based approaches have been
proposed~\cite{tamar2016value,karkus2017qmdp}.
For example, a neural network with a model predictive control module~\cite{amos2018differentiable,agrawal2020learning}
and a neural network with a linear quadratic regulator module in a Koopman invariant subspace~\cite{iwata2021controlling} have been used
for controllers, or policy networks.
However, these methods cannot be used to control the frequency and convergence rate.

\begin{figure}[t!]
  \centering
  \includegraphics[width=26em]{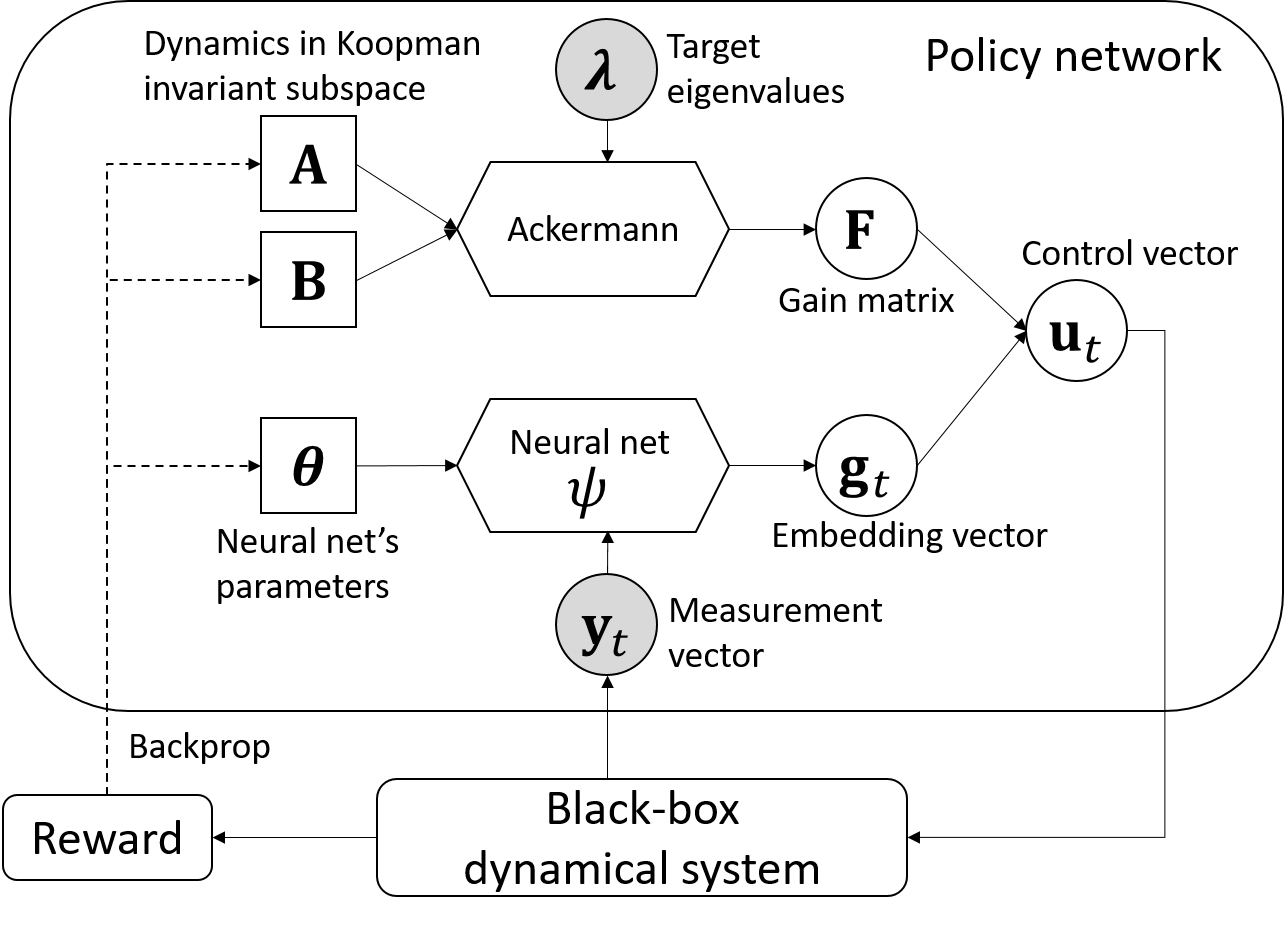}
  \caption{Our proposed method.
    In the policy network, shaded circles represent given variables, unshaded circles represent inferred variables,
    squares represent parameters to be trained, and hexagons represent functions.
    The policy network takes measurement vector $\vec{y}_{t}$
    as input from the black-box dynamical system. The measurement vector is transformed to Koopman embedding $\vec{g}_{t}$
    by neural network $\psi$ with parameters $\bm{\theta}$.
    Gain matrix $\vec{F}$ is obtained by the Ackermann's method using
    target eigenvalues $\bm{\lambda}$ and dynamics parameters $\vec{A}$ and $\vec{B}$ in the Koopman invariant subspace.
    Control vector $\vec{u}_{t}$ is calculated using $\vec{g}_{t}$ and $\vec{F}$,
    and it is passed to the black-box dynamical system.
    The reward is calculated using the difference between the target eigenvalues
    and eigenvalues of the controlled dynamics.
    The reward is backpropagated to update parameters of the policy network $\vec{A}$, $\vec{B}$, and $\bm{\theta}$ to maximize the expected reward.}
  \label{fig:model}
\end{figure}

We propose an end-to-end learning method for controlling black-box nonlinear dynamical systems
to have a desired frequency and convergence rate using reinforcement learning
by modeling dynamics in a Koopman invariant subspace.
We are given target eigenvalues of a Koopman operator that specify the desired frequency and convergence rate.
The proposed method trains a policy network
such that the eigenvalues of the Koopman operator of the controlled dynamics are the same with the target eigenvalues.
Our policy network consists of an embedding function based on a neural network,
and a pole placement module.
The embedding function is used to find an appropriate Koopman invariant subspace.
The pole placement module is used to find an optimal controller
in the Koopman invariant subspace based on the Ackermann's method~\cite{ackermann2009pole}.
Since the pole placement module is differentiable,
the neural network and Koopman dynamics
can be trained in an end-to-end fashion
by directly minimizing the difference between the target eigenvalues and eigenvalues of the controlled dynamics
using the policy gradient method~\cite{sutton1999policy}.
The proposed method exploits the advantages of model-free and model-based approaches
by incorporating the pole placement module in a neural network based on the Koopman operator theory.
Figure~\ref{fig:model} illustrates our proposed method, which is explained in detail in Section~\ref{sec:proposed}.

\section{Preliminaries: Koopman operator theory}
\label{sec:preliminaries}

We briefly review the Koopman operator theory in this section.
We consider nonlinear discrete-time dynamical system,
$\vec{x}_{t+1}=f(\vec{x}_{t})$,
where $\vec{x}_{t}\in\mathcal{X}$ is the state at timestep $t$.
Koopman operator $\mathcal{A}$ is defined as an infinite-dimensional linear operator that acts on observables $g:\mathcal{X}\rightarrow\mathbb{R}$ (or $\mathbb{C}$)~\cite{koopman1931hamiltonian},
$g(\vec{x}_{t+1})=\mathcal{A}g(\vec{x}_{t})$,
with which the analysis of nonlinear dynamics
can be lifted to a linear (but infinite-dimensional) regime.
The eigenvalues $\{\lambda_{k}\}$ of the Koopman operator characterize the time evolution.
In particular, its argument $\arg \lambda_{k}$ determines the frequency, and
its absolute value $|\lambda_{k}|$ determines the convergence rate.
Although the existence of the Koopman operator is theoretically guaranteed in various situations, its practical use is limited by its infinite dimensionality. 
We can assume the restriction of $\mathcal{A}$ 
to a finite-dimensional subspace,
which results in finite-dimensional operator, 
$\vec{g}_{t+1}=\vec{A}\vec{g}_{t}$,
where
$\vec{A}\in\mathbb{R}^{K\times K}$ is a finite-dimensional approximation of the Koopman operator,
and
$\vec{g}_{t}=[g_{1}(\vec{x}_{t}),\dots,g_{K}(\vec{x}_{t})]\in\mathbb{R}^{K}$ is an embedding vector
of state $\vec{x}_{t}$ in the Koopman invariant subspace at timestep $t$.

\section{Proposed method}
\label{sec:proposed}

\subsection{Problem formulation}

We are given $K$ target eigenvalues $\bm{\lambda}=\{\lambda_{k}\}_{k=1}^{K}$,
where $\lambda_{k}\in\mathbb{C}$ is the $k$th target eigenvalue.
We can access a black-box nonlinear dynamical system
\begin{align}
    \vec{x}_{t+1}=f(\vec{x}_{t},\vec{u}_{t}),\quad
    \vec{y}_{t}=h(\vec{x}_{t}),
    \label{eq:dynamics}
\end{align}
where $\vec{x}_{t}\in\mathcal{X}$ is the state,
$\vec{y}_{t}\in\mathbb{R}^{D}$ is the measurement vector, 
$\vec{u}_{t}\in\mathbb{R}^{J}$ is the control vector at timestep $t$.
We do not know functions $f$ and $h$, 
but we can observe measurement vector $\vec{y}_{t}$ 
and select control vector $\vec{u}_{t}$ for each timestep.
Our aim is to find control sequence $(\vec{u}_{t})_{t=1}^{T}$
where eigenvalues $\hat{\bm{\lambda}}$ of the Koopman operator
of the controlled dynamics are close to target eigenvalues $\bm{\lambda}$.

\subsection{Policy network}

In this subsection, we describe our policy network
that outputs control vector $\vec{u}$ given measurement vector $\vec{y}$.
We embed measurement vectors into a Koopman invariant space by a neural network,
which enables us to apply pole placement techniques for linear dynamics to nonlinear dynamics,
\begin{align}
  \vec{g}_{t}=\psi(\vec{y}_{t};\bm{\theta}),
  \label{eq:g}
\end{align}
where 
$\psi:\mathbb{R}^{D}\rightarrow\mathbb{R}^{K}$ is a neural network,
and $\bm{\theta}$ is its parameters.
In the Koopman invariant subspace,
embedding vectors are assumed to obey the following linear dynamics with control~\cite{proctor2016dynamic,BruderGRV19,korda2020optimal},
\begin{align}
  \vec{g}_{t+1}=\vec{A}\vec{g}_{t}+\vec{B}\vec{u}_{t},
  \label{eq:g_dynamics}
\end{align}
where $\vec{A}\in\mathbb{R}^{K\times K}$ is a finite-dimensional approximation of the Koopman operator,
and $\vec{B}\in\mathbb{R}^{K\times L}$ is the linear effect of a control vector on the embedding vector
at the next timestep.

We can obtain optimal control vector $\vec{u}_{t}$
by a linear projection of embedding vector $\vec{g}_{t}$,
\begin{align}
  \vec{u}_{t}=-\vec{F}\vec{g}_{t},
  \label{eq:u}
\end{align}
where $\vec{F}\in\mathbb{R}^{L\times K}$ is a gain matrix.
Then, controlled dynamics in the Koopman invariant subspace is given by
\begin{align}
  \vec{g}_{t+1}=(\vec{A}-\vec{B}\vec{F})\vec{g}_{t}.
  \label{eq:g_control}
\end{align}
It is desirable that the eigenvalues of $\vec{A}-\vec{B}\vec{F}$
are the same with target eigenvalues $\bm{\lambda}$.
We achieve this by obtaining gain matrix $\vec{F}$
using the Ackermann's method as follows,
\begin{align}
  \vec{F}=[0\quad0\quad\cdots\quad1]\vec{C}^{-1}\Delta(\vec{A};\bm{\lambda}),
  \label{eq:f}
\end{align}
where
$\vec{C}$ is a controllability matrix defined by
\begin{align}
  \label{eq:c}
\vec{C}=[\vec{B}\quad \vec{A}\vec{B}\quad\cdots\quad\vec{A}^{K-1}\vec{B}],
\end{align}
$\Delta(\vec{A};\bm{\lambda})$ is the desired characteristic polynomial evaluated at $\vec{A}$,
\begin{align}
  \Delta(\vec{A};\bm{\lambda})=\vec{A}^{K}+\sum_{k=1}^{K}\beta_{k}\vec{A}^{K-k},
  \label{eq:delta}
\end{align}
and $\beta_{k}$ is the coefficients of a polynomial with
target eigenvalues $\bm{\lambda}$ of roots,
\begin{align}
  \prod_{k=1}^{K}(s-\lambda_{k})=s^{K}+\sum_{k=1}^{K}\beta_{k}r^{K-k}.
\end{align}
Using Eqs.~(\ref{eq:g},\ref{eq:u},\ref{eq:f}),
policy network $\pi$, which outputs control vector $\vec{u}$
given measurement vector $\vec{y}$, is written as follows,
\begin{align}
  \vec{u}=\pi(\vec{y};\bm{\Theta})\equiv-[0\quad0\quad\cdots\quad1]
  \vec{C}^{-1}\Delta(\vec{A};\bm{\lambda})\psi(\vec{y};\bm{\theta}),
  \label{eq:pi}
\end{align}
where $\bm{\Theta}=\{\vec{A},\vec{B},\bm{\theta}\}$ is parameters of the policy network to be trained.
The policy network is differentiable with respect to the parameters.
Unlike policy networks in the existing reinforcement learning methods,
our policy network incorporates the Ackermann's method and dynamics in the Koopman invariant subspace as well as neural networks,
which enables us to use the knowledge of well-studied pole placement control with reinforcement learning.

Our policy network assumes linear dynamics with control in Eq.~(\ref{eq:g_dynamics}),
and its controllability in the Koopman invariant subspace.
Since any controlled system of form $\vec{x}_{t+1}=f'(\vec{x}_{t},\vec{u}')$
can be transformed to $\vec{x}_{t+1}=f''(\vec{x}_{t})+\vec{B}'\vec{u}''$ by
the state inflation~\cite{korda2020optimal},
we can assume Eq.~(\ref{eq:g_dynamics}) when $f''$ is approximated well by the dynamics
in a Koopman invariant subspace.
Even when these assumptions are not met perfectly,
the proposed method tries to find an appropriate Koopman invariant subspace
where our policy network works 
by flexibly modifying the neural network
with end-to-end training that directly improves the control performance.

\subsection{Training}

We train the parameters of the policy network $\bm{\Theta}$
using reinforcement learning,
where a state is measurement vector $\vec{y}$,
an action is control vector $\vec{u}$,
and a negative reward is the distance between target eigenvalues
$\bm{\lambda}$ and
eigenvalues $\hat{\bm{\lambda}}=\{\hat{\lambda}_{k}\}_{k=1}^{K}$
of the Koopman operator of the controlled dynamics.
Specifically, the reward is defined by
\begin{align}
  r = -\frac{1}{K}\sum_{k}\min_{k'}|\lambda_{k}-\hat{\lambda}_{k'}|.
  \label{eq:r}
\end{align}
The procedures to estimate eigenvalues $\hat{\bm{\lambda}}$
is described in Section~\ref{sec:eigenvalue}.

The objective function to be maximized is the expected cumulative reward,
\begin{align}
  J(\bm{\Theta})=\mathbb{E}_{\vec{Y}|\bm{\Theta}}\left[\sum_{t=1}^{T}\gamma^{t-1}r(\vec{Y})\right],
  \label{eq:j}
\end{align}
where $\mathbb{E}_{\vec{Y}|\bm{\Theta}}$ is the expectation over
measurement sequences when controlled
by the policy network with parameters $\bm{\Theta}$,
$\gamma\in(0,1]$ is discount factor,
and $r_{t}(\vec{Y})$ is the reward at timestep $t$
with measurement sequence $\vec{Y}=(\vec{y}_{1},\cdots,\vec{y}_{T})$
calculated by Eq.~(\ref{eq:r}).
The control vectors to be used next are sampled from
the following Gaussian distribution for exploration,
\begin{align}
  p(\vec{u}|\vec{y};\bm{\Theta})=\mathcal{N}(\pi(\vec{y};\bm{\Theta}),\sigma^{2}\vec{I}).
  \label{eq:pu}
\end{align}
The objective function in Eq.~(\ref{eq:j})
is maximized by the policy gradient method~\cite{sutton1999policy}.

The training procedure is shown in Algorithm~\ref{alg}.
The expectation in Eq.~(\ref{eq:j}) is approximated by the Monte Carlo method
using sampled measurement vectors, control vectors, and cumulative discount rewards.
We use the average of the discounted sum of the future rewards
as a baseline to reduce the
variance~\cite{weaver2001optimal}.
Although the reward is given at last timestep $T$ in our experiments,
it can be given at all timesteps.
We can add other terms in the reward,
such as control cost $\sum_{t=1}^{T}\parallel\vec{u}_{t}\parallel^{2}$.

\begin{algorithm}[t!]
  \caption{Training procedure.}
  \label{alg}
  \begin{algorithmic}[1]
    \renewcommand{\algorithmicrequire}{\textbf{Input:}}
    \renewcommand{\algorithmicensure}{\textbf{Output:}}
    \REQUIRE{Black-box dynamical system, 
    target eigenvalues $\bm{\lambda}$,
    control variance $\sigma^{2}$,
    discount factor $\gamma$.}
    \ENSURE{Trained parameters $\bm{\Theta}$.}
    \WHILE{End condition is satisfied}
    \STATE Calculate optimal gain matrix $\vec{F}$ by Eq.~(\ref{eq:f}) using current parameters $\bm{\Theta}$.
    \STATE Randomly sample initial measurement vector $\vec{y}_{1}$.
    \FOR{$t \in \{1,\cdots,T-1\}$}
    \STATE Calculate the mean of the control distribution $\pi(\vec{y}_{t};\bm{\Theta})$ by Eq.~(\ref{eq:pi}).
    \STATE Sample control vector $\vec{u}_{t}$ according to the control distribution in Eq.~(\ref{eq:pu}).
    \STATE Obtain next measurement vector $\vec{y}_{t+1}$ from the system with sampled control $\vec{u}_{t}$.
    \ENDFOR
    \STATE Estimate eigenvalues $\hat{\bm{\lambda}}$ of the Koopman operator of measurement sequence $\vec{Y}$ as described in Section~\ref{sec:eigenvalue}.
    \STATE Calculate reward $r$ by Eq.~(\ref{eq:r}).
    \STATE Calculate cumulative discounted rewards $R_{t}(\vec{Y})$ for $t\in\{1,\cdots,T\}$.
    \STATE Calculate objective function $J$ in Eq.~(\ref{eq:j}) and its gradient.
    \STATE Update model parameters $\bm{\Theta}$ using $J$ and its gradient using a stochastic gradient method.
    \ENDWHILE
  \end{algorithmic}
\end{algorithm}

\subsection{Eigenvalue estimation}
\label{sec:eigenvalue}

We estimate eigenvalues $\hat{\bm{\lambda}}$ of the controlled sequence
by dynamic mode decomposition (DMD) on time delay coordinates,
or Hankel DMD~\cite{arbabi2017ergodic},
which can theoretically yield the eigenvalues of a Koopman operator.
Let $\vec{Y}=(\vec{y}_{1},\cdots,\vec{y}_{T})$
be a sequence of measurement vectors with length $T$
that are obtained by the given black-box dynamical system controlled by the policy network.
Let $\vec{H}_{1},\vec{H}_{2}\in\mathbb{R}^{D\tau\times(T-\tau)}$ be Hankel matrices with time delay $\tau$,
where $\vec{H}_{1}$ contains $\vec{y}_{1}$ to $\vec{y}_{T-1}$,
and $\vec{H}_{2}$ contains $\vec{y}_{2}$ to $\vec{y}_{T}$.
The Koopman operator is approximated by
\begin{align}
  \hat{\vec{A}}=\vec{U}^{\top}\vec{H}_{2}\vec{V}\bm{\Sigma}^{-1}\in\mathbb{R}^{K\times K},
\end{align}
where $\vec{U}$, $\vec{V}$, and $\bm{\Sigma}$ 
are obtained by the singular value decomposition of $\vec{H}_{1}$
with rank $K$,
$\vec{H}_{1}\approx\vec{U}\bm{\Sigma}\vec{V}^{\top}$,
$\vec{U}\in\mathbb{R}^{D\tau\times K}$,
$\bm{\Sigma}\in\mathbb{R}^{K\times K}$, and
$\vec{U}\in\mathbb{R}^{(T-\tau)\times K}$.
By the low-rank approximation, we can reduce the noise
in the measurement vectors for modeling the dynamics
in the Koopman invariant subspace.
The estimated eigenvalues are obtained by the eigen decomposition of
the approximated Koopman operator,
\begin{align}
  \hat{\vec{A}}=\bm{\Phi}\mathrm{diag}([\hat{\lambda}_{1},\cdots,\hat{\lambda}_{K}])\bm{\Phi}^{-1},
\end{align}
where $\vec{\Phi}\in\mathbb{C}^{K\times K}$ is linearly independent eigenvectors.

\section{Experiments}

\subsection{Dynamical systems}

\begin{figure}[t!]
  \centering
  {\tabcolsep=0.3em
    \begin{tabular}{cccc}
  Van der Pol & Fitzhugh-Nagumo & Duffing & Rossler\\
  \includegraphics[width=10.8em]{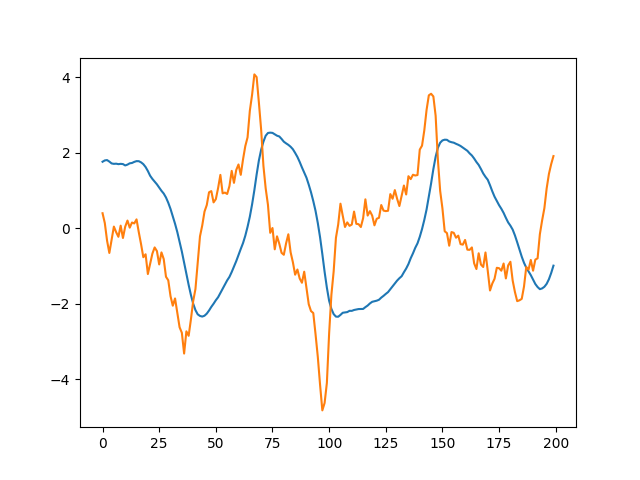}&  
  \includegraphics[width=10.8em]{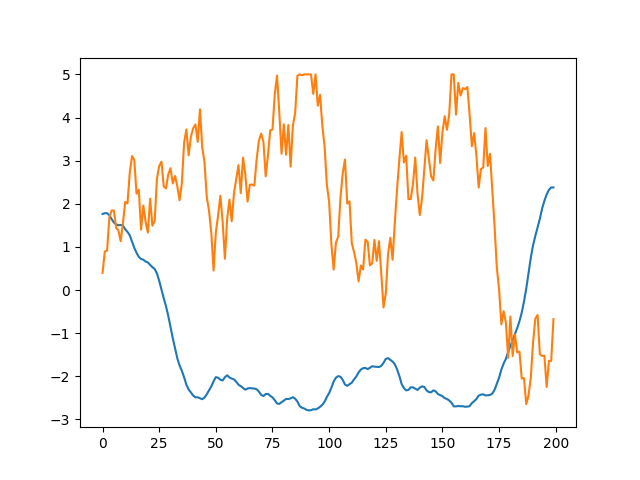}&
  \includegraphics[width=10.8em]{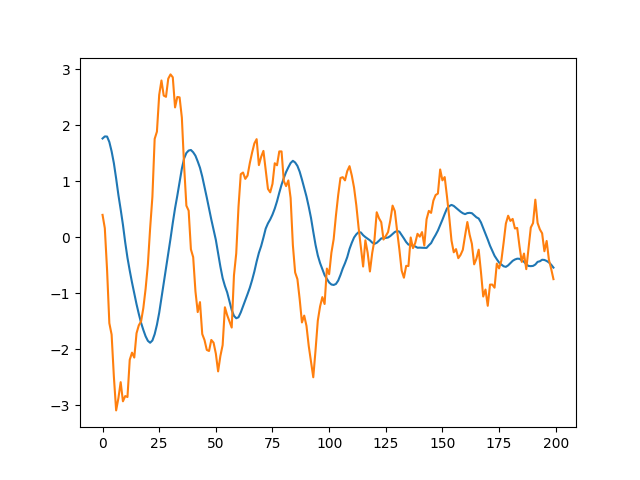}&
  \includegraphics[width=10.8em]{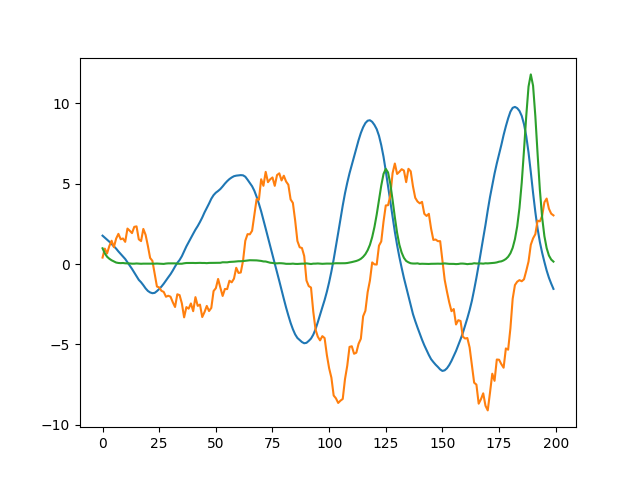}\\
  \end{tabular}}
  \caption{Measurement vector sequences with random control.
      The horizontal axis is time, and the vertical axis is values for each element of the measurement vectors.}
\label{fig:data}
\end{figure}

To evaluate the proposed method,
we used the following four nonlinear dynamical systems:
Van der Pol, Fitzhugh-Nagumo, Duffing, and Rossler,
where we added scalar control variable $u_{1}$ to the original systems.
Figure~\ref{fig:data} shows
examples of measurement vector sequences
with random control.
For all systems,
measurement vectors were obtained by $\vec{y}_{t}=\vec{x}_{t}+\epsilon$,
where $\epsilon$ was Gaussian noise with
mean $\bm{0}$ and standard deviation $10^{-2}$.

The Van der Pol equation is a non-conservative oscillator with nonlinear
dampling~\cite{van1927frequency}.
We used the following
Van der Pol system with control,
$\frac{dx_{1}}{dt}=x_{2},\frac{dx_{2}}{dt}=ax_{2}(1-x_{1}^{2})-bx_{1}+u_{1}$,
where $a=1$ and $b=1$.
The state and control variables were bound in
$x_{d}\in[-10,10]$ and $u_{1}\in[-5,5]$.

The Fitzhugh-Nagumo equation~\cite{fitzhugh1955mathematical,nagumo1962active} is a model
of an excitable system such as a neuron.
We used the following FitzHugh-Nagumo system with control,
$\frac{dx_{1}}{dt}=x_{1}-\frac{x_{1}^{3}}{3}-x_{2}+I,
\frac{dx_{2}}{dt}=c(x_{1}-a-bx_{2})+u_{1}$,
where $a=0.7$, $b=0.8$, $c=0.08$, and $I=0.8$.
The state and control variables were bound in
$x_{d}\in[-10,10]$ and $u_{1}\in[-5,5]$.

The Duffing system is a non-linear second-order differential equation
used to model damped and driven oscillators~\cite{guckenheimer2013nonlinear},
We used the following Duffing system with control,
$\frac{dx_{1}}{dt}=x_{2},
\frac{dx_{2}}{dt}=b x_{1}-a x_{1}^{3}-c x_{2}+u_{1}$,
where $a=1$, $b=-1$, and $c=0.5$. 
The state and control variables were bound in
$x_{d}\in[-5,5]$ and $u_{1}\in[-10,10]$.

The Rossler system is a non-linear differential equation~\cite{rossler1976equation} that exhibits chaotic dynamics.
We used the following Rossler system with control,
$\frac{dx_{1}}{dt}=-x_{2}-x_{1},
\frac{dx_{2}}{dt}=x_{1}-ax_{2}+u_{1},
\frac{dx_{3}}{dt}=b+x_{1}x{3}-cx_{3}$,
where $a=0.2$, $b=0.2$, and $c=5.7$.
The state and control variables were bound in
$x_{d}\in[-20,20]$ and $u_{1}\in[-10,10]$.

\subsection{Comparing methods}

We compared the proposed method with the following three methods:
system identification with a linear model (SL),
system identification with a nonlinear model (SN),
and model-free reinforcement learning (RL),
where SL and SN are two-step approaches,
and RL and the proposed method is end-to-end approaches.

With SL,
the system is firstly identified by assuming
linear dynamical model $\vec{y}_{t+1}=\vec{A}\vec{y}_{t}+\vec{B}\vec{u}_{t}$.
Then, optimal optimal gain matrix $\vec{F}$ is obtained by Eq.~(\ref{eq:f})
using estimated parameters $\vec{A}$ and $\vec{B}$
based on pole placement while fixing the identified system.
The parameters $\vec{A}$ and $\vec{B}$ are estimated
using a set of measurement and control sequences
$\mathcal{D}=\{((\vec{y}_{nt},\vec{u}_{nt}))_{t=1}^{T}\}_{n=1}^{N}$,
where $\vec{y}_{nt}$ and $\vec{u}_{nt}$ are the $t$th measurement
and control vectors in the $n$th sequence.
The measurement vectors were obtained
by the black-box dynamical system,
and the control vectors were obtained uniform randomly.

With SN,
the system is firstly identified by assuming
linear dynamical model in a Koopman invariant subspace
$\psi(\vec{y}_{t+1};\bm{\theta})=\vec{A}\psi(\vec{y}_{t};\bm{\theta})+\vec{B}\vec{u}_{t}$
using a set of sequences $\mathcal{D}$,
where $\psi$ is an encoder neural network with parameters $\bm{\theta}$.
Then, optimal optimal gain matrix $\vec{F}$ is obtained as with SL while fixing the identified system.
The parameters of the linear dynamical model and encoder neural network
are trained by minimizing
the following sum of the reconstruction and prediction errors~\cite{takeishi2017learning}.

With RL,
policy network $\pi$ is modeled by a feed-forward neural network that takes
measurement vector $\vec{y}$ as input, and outputs control vector $\vec{u}$.
Unlike the proposed method,
the policy network of RL does not contain the pole placement module
in the Koopman invariant subspace.
The parameters of the neural network are trained by the policy gradient
as with the proposed method.

\subsection{Settings}

In the proposed method,
we used a three-layered feed-forward neural network
with eight hidden units and two output units.
The dimensionality of the Koopman invariant subspace was $K=2$.
The activation function in the neural networks
was rectified linear unit, $\mathrm{ReLU}(x)=\max(0,x)$.
Optimization was performed using Adam~\cite{kingma2014adam}
with learning rate $10^{-3}$ and batch size ten.
The discount factor was $\gamma=0.99$.
The variance of control $\sigma^{2}$ was set to half of
the bounded width of the control variable.
The maximum number of training epochs was 10,000,
and the validation simulation runs were used for early stopping.
The number of timesteps for each sequence was $T=200$.
For estimating the eigenvalues of control dynamics,
we used Hankel DMD with five-timestep time delay. 
In SN, we used a neural network with the same architecture with the proposed method for encoder $g$,
and three-layered feed-forward neural network with eight hidden units for
decoder $g'$.
In RL, we used a four-layered feed-forward neural network
with eight and two hidden units and one output unit.
In SL and SN, we used 100 measurement and control vector sequences obtained
with random control,
and we used long-term prediction errors with $L=4$ in the objective function.
In the proposed method, the parameters were pretrained using SN.
In all methods, when output control vectors were outside of the bound,
they were clipped to the bound.

\subsection{Results}

\begin{table}[t!]
  \centering
  \caption{Average mean absolute errors of eigenvalues and its standard errors.
    $|\lambda|$ and $\arg \lambda$ shows the absolute value and argument of the target eigenvalue, respectively.
    Values in bold are not statistically different at 5\% level from the best performing method in each data by a paired t-test.}
  \label{tab:error}
  (a) Van der Pol \\
  {\tabcolsep=0.4em
  \begin{tabular}{rrrrrr}
\hline    
$|\lambda|$ & $\arg \lambda$ & Ours & SL & SN & RL \\
\hline
1.00& 0.0 & 0.005$\pm$0.001 & 0.017$\pm$0.001 & 0.030$\pm$0.005 & {\bf 0.002$\pm$0.001}\\
1.00& 0.1 & {\bf 0.004$\pm$0.000} & 0.005$\pm$0.000 & 0.007$\pm$0.001 & 0.120$\pm$0.022\\
1.00& 0.2 & {\bf 0.004$\pm$0.000} & 0.009$\pm$0.001 & 0.020$\pm$0.002 & 0.253$\pm$0.046\\
1.00& 0.3 & {\bf 0.068$\pm$0.010} & 0.187$\pm$0.000 & 0.199$\pm$0.001 & 0.435$\pm$0.051\\
 0.96 & 0.2 & {\bf 0.057$\pm$0.006} & 0.080$\pm$0.000 & 0.079$\pm$0.001 & 0.253$\pm$0.050\\
 0.92 & 0.2 & {\bf 0.075$\pm$0.009} & 0.126$\pm$0.002 & 0.102$\pm$0.004 & 0.290$\pm$0.043\\
 \hline
  \end{tabular}}
  
  (b) Fitzhugh-Nagumo \\
  {\tabcolsep=0.4em
  \begin{tabular}{rrrrrr}
\hline    
$|\lambda|$ & $\arg \lambda$ & Ours & SL & SN & RL \\
\hline  
  1.00& 0.0 & {\bf 0.003$\pm$0.000} & 0.018$\pm$0.000 & 0.011$\pm$0.001 & {\bf 0.003$\pm$0.001}\\
1.00& 0.1 & {\bf 0.003$\pm$0.000} & 0.120$\pm$0.005 & 0.078$\pm$0.005 & 0.183$\pm$0.009\\
1.00& 0.2 & {\bf 0.002$\pm$0.000} & 0.060$\pm$0.001 & 0.075$\pm$0.002 & 0.332$\pm$0.033\\
1.00& 0.3 & {\bf 0.008$\pm$0.001} & 0.115$\pm$0.001 & 0.076$\pm$0.003 & 0.504$\pm$0.051\\
 0.96 & 0.2 & {\bf 0.038$\pm$0.008} & 0.100$\pm$0.001 & 0.136$\pm$0.003 & 0.343$\pm$0.036\\
 0.92 & 0.2 & {\bf 0.107$\pm$0.021} & 0.172$\pm$0.001 & 0.175$\pm$0.010 & 0.362$\pm$0.030\\
 \hline
  \end{tabular}}
  
  (c) Duffing \\
  {\tabcolsep=0.4em
  \begin{tabular}{rrrrrr}
\hline    
$|\lambda|$ & $\arg \lambda$ & Ours & SL & SN & RL \\
\hline  
1.00& 0.0 & 0.008$\pm$0.001 & 0.195$\pm$0.005 & 0.146$\pm$0.007 & {\bf 0.004$\pm$0.002}\\
1.00& 0.1 & {\bf 0.042$\pm$0.007} & 0.074$\pm$0.003 & 0.058$\pm$0.002 & 0.149$\pm$0.016\\
1.00& 0.2 & {\bf 0.019$\pm$0.001} & 0.039$\pm$0.002 & 0.041$\pm$0.001 & 0.304$\pm$0.036\\
1.00& 0.3 & {\bf 0.006$\pm$0.001} & 0.040$\pm$0.001 & 0.046$\pm$0.002 & 0.476$\pm$0.052\\
 0.96 & 0.2 & {\bf 0.031$\pm$0.003} & 0.039$\pm$0.001 & 0.044$\pm$0.001 & 0.305$\pm$0.042\\
 0.92 & 0.2 & 0.060$\pm$0.007 & {\bf 0.042$\pm$0.001} & 0.046$\pm$0.002 & 0.310$\pm$0.044\\
 \hline
  \end{tabular}}
  
  (d) Rossler \\
  {\tabcolsep=0.4em
  \begin{tabular}{rrrrrr}
\hline    
$|\lambda|$ & $\arg \lambda$ & Ours & SL & SN & RL \\
\hline   
1.00& 0.0 & {\bf 0.001$\pm$0.000} & 0.121$\pm$0.001 & 0.044$\pm$0.006 & 0.008$\pm$0.003\\
1.00& 0.1 & {\bf 0.022$\pm$0.006} & 0.064$\pm$0.001 & 0.062$\pm$0.017 & 0.103$\pm$0.017\\
1.00& 0.2 & {\bf 0.032$\pm$0.006} & 0.257$\pm$0.001 & 0.111$\pm$0.040 & 0.224$\pm$0.034\\
1.00& 0.3 & {\bf 0.141$\pm$0.046} & 0.456$\pm$0.001 & 0.282$\pm$0.052 & 0.398$\pm$0.040\\
 0.96 & 0.2 & {\bf 0.081$\pm$0.028} & 0.296$\pm$0.008 & {\bf 0.139$\pm$0.039} & 0.213$\pm$0.039\\
 0.92 & 0.2 & {\bf 0.122$\pm$0.019} & 0.382$\pm$0.012 & {\bf 0.179$\pm$0.037} & 0.261$\pm$0.035\\
\hline
  \end{tabular}}
\end{table}

\begin{figure*}[t!]
  \centering
  {\tabcolsep=-0.35em
    \begin{tabular}{ccccc}
      \multicolumn{5}{c}{Van der Pol}\vspace{-0.1em}\\
      \includegraphics[width=10.0em]{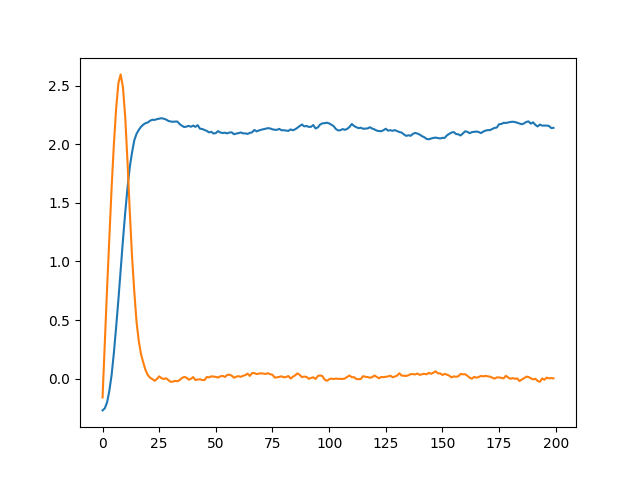}&
      \includegraphics[width=10.0em]{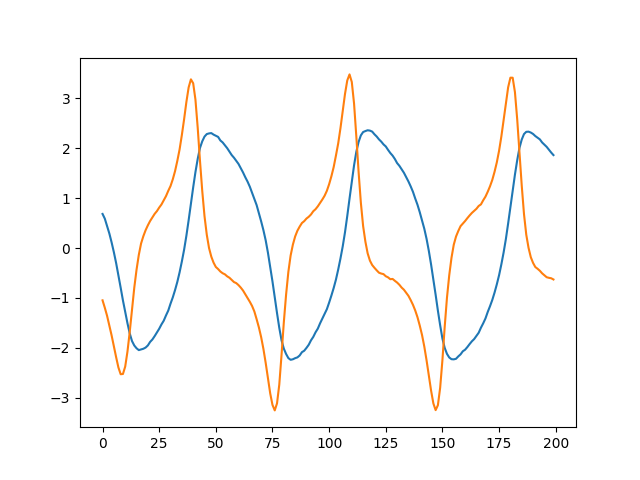}&
      \includegraphics[width=10.0em]{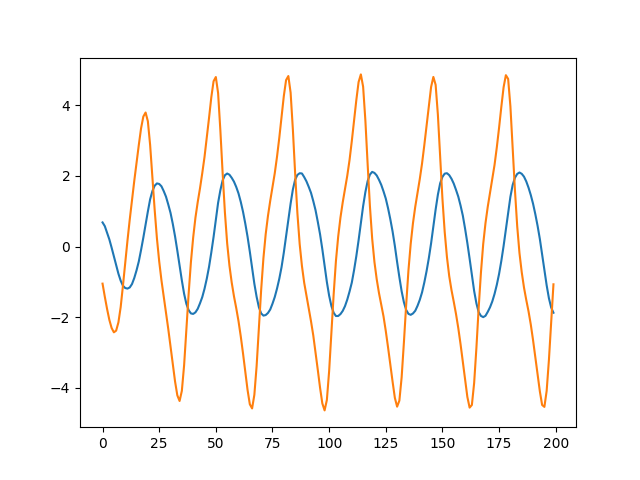}&
      \includegraphics[width=10.0em]{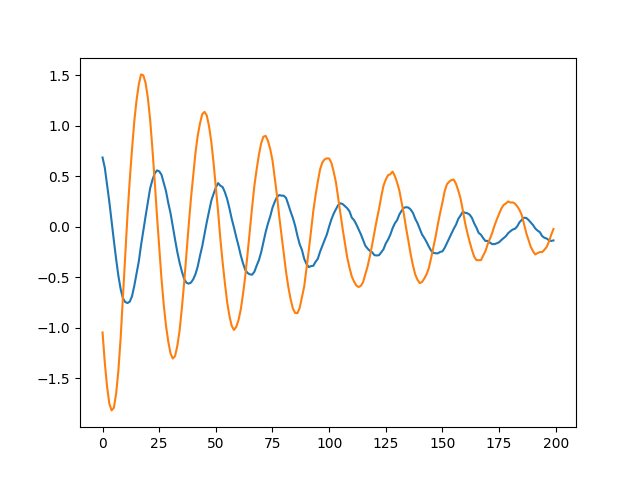}&
      \includegraphics[width=10.0em]{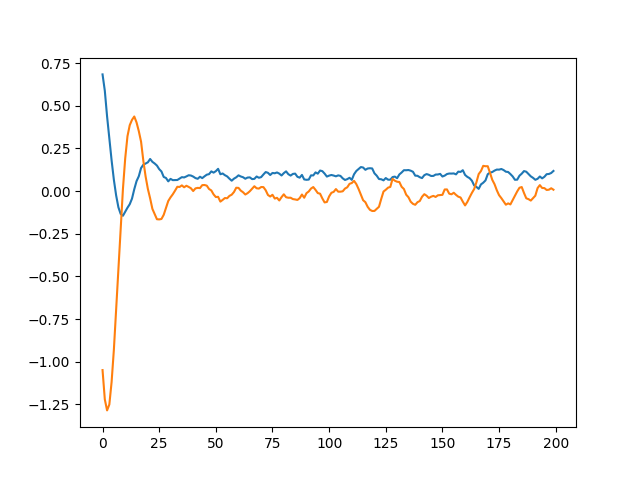}\\
      \multicolumn{5}{c}{Fitzhugh-Nagumo}\vspace{-0.1em}\\
      \includegraphics[width=10.0em]{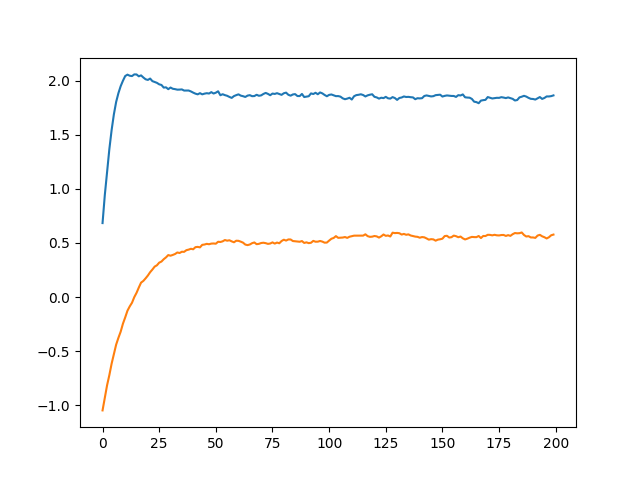}&
      \includegraphics[width=10.0em]{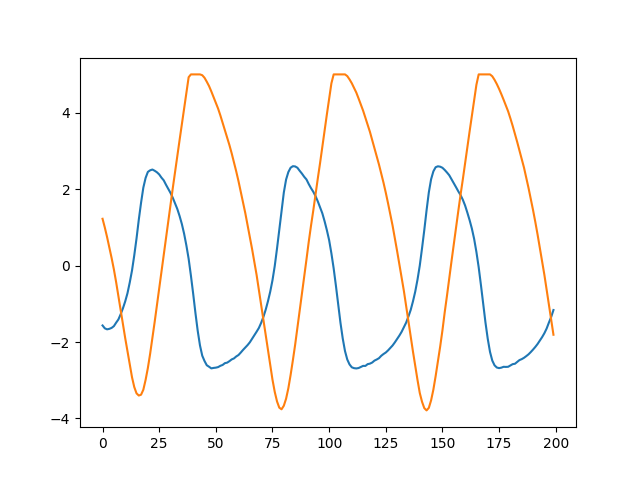}&
      \includegraphics[width=10.0em]{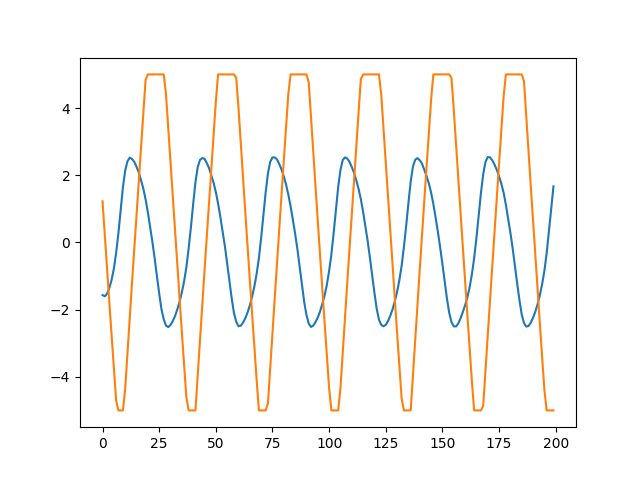}&
      \includegraphics[width=10.0em]{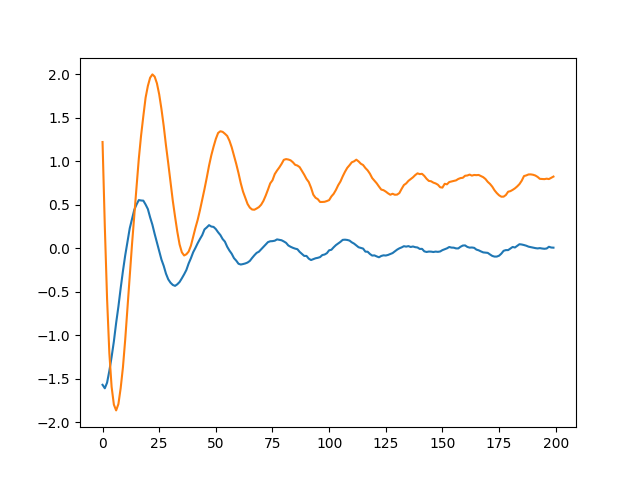}&
      \includegraphics[width=10.0em]{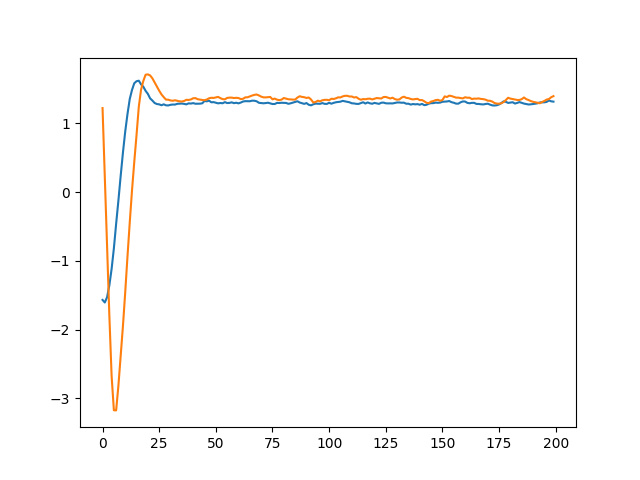}\\
      \multicolumn{5}{c}{Duffing}\vspace{-0.1em}\\
      \includegraphics[width=10.0em]{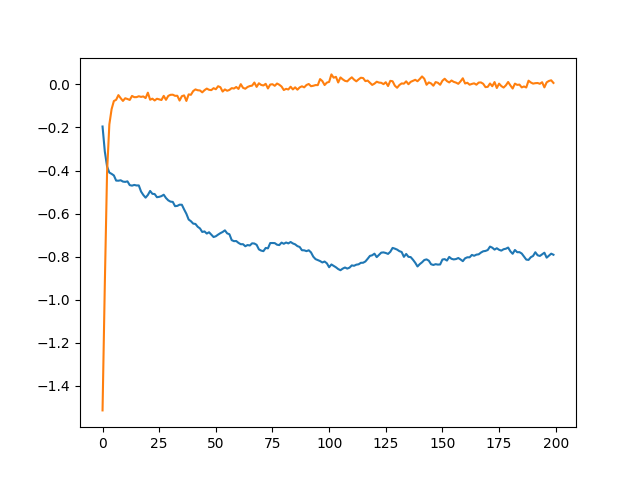}&
      \includegraphics[width=10.0em]{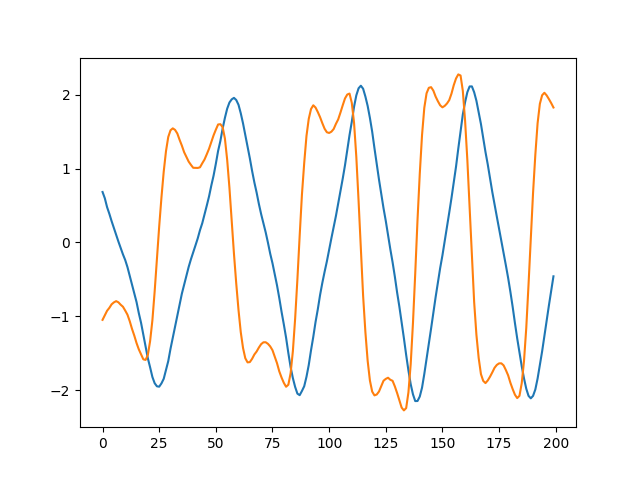}&
      \includegraphics[width=10.0em]{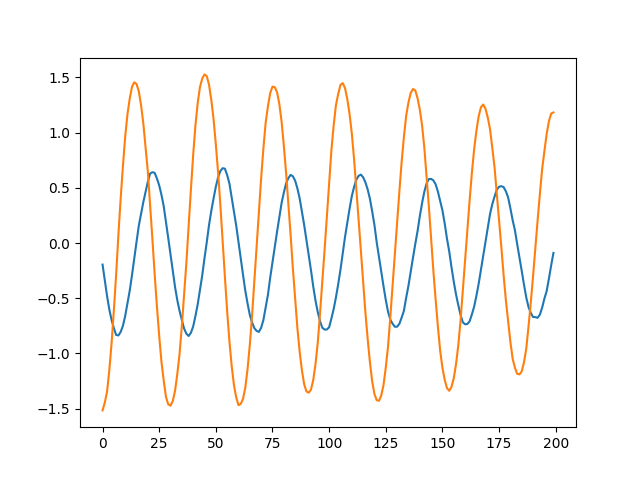}&
      \includegraphics[width=10.0em]{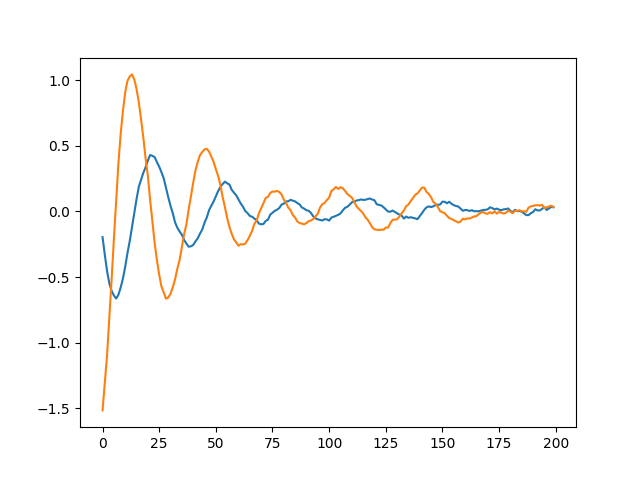}&
      \includegraphics[width=10.0em]{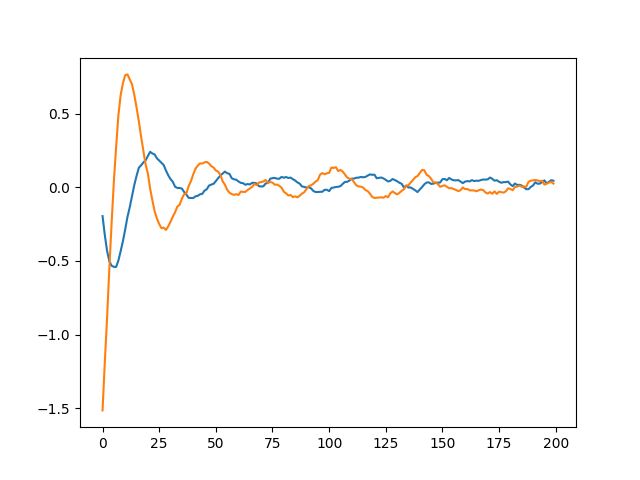}\\
      \multicolumn{5}{c}{Rossler}\vspace{-0.1em}\\
      \includegraphics[width=10.0em]{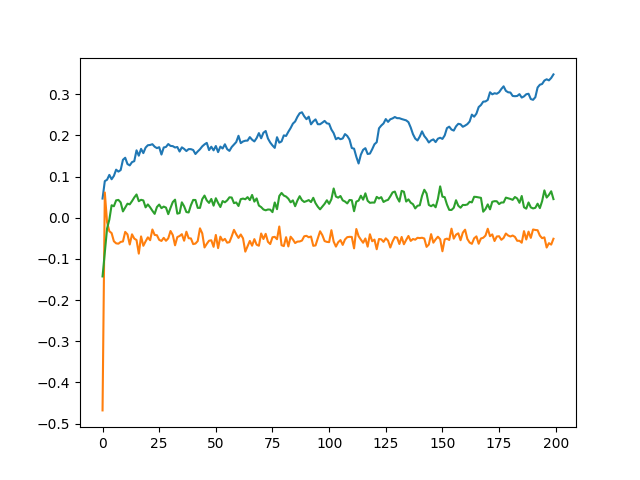}&
      \includegraphics[width=10.0em]{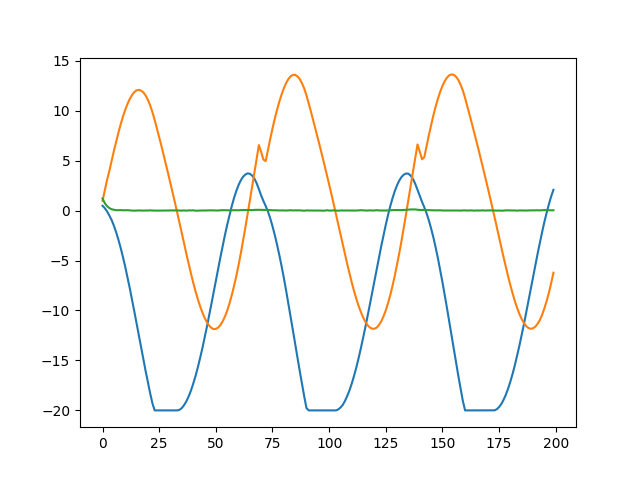}&
      \includegraphics[width=10.0em]{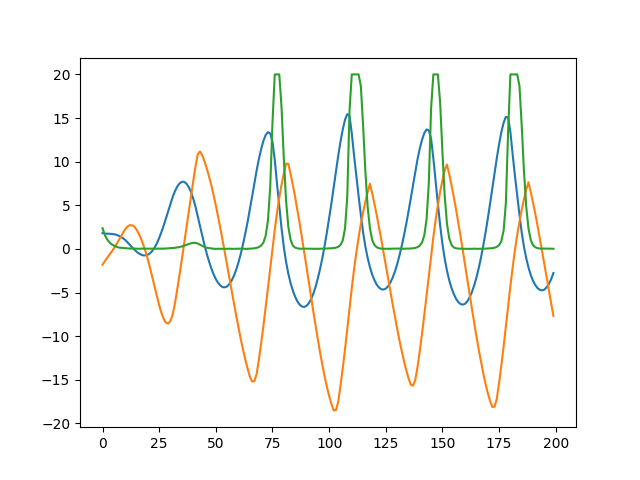}&
      \includegraphics[width=10.0em]{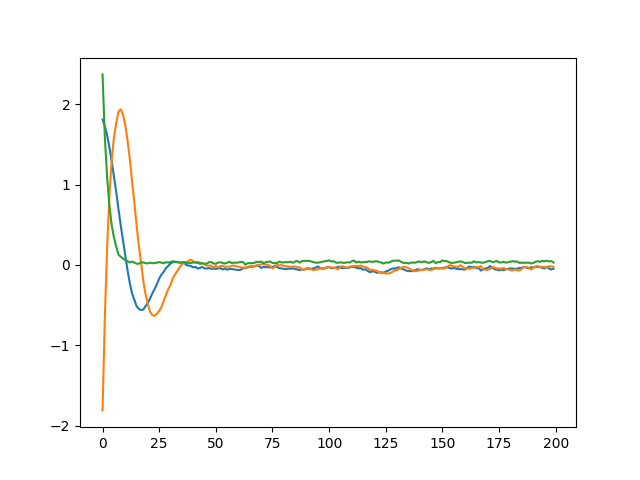}&
      \includegraphics[width=10.0em]{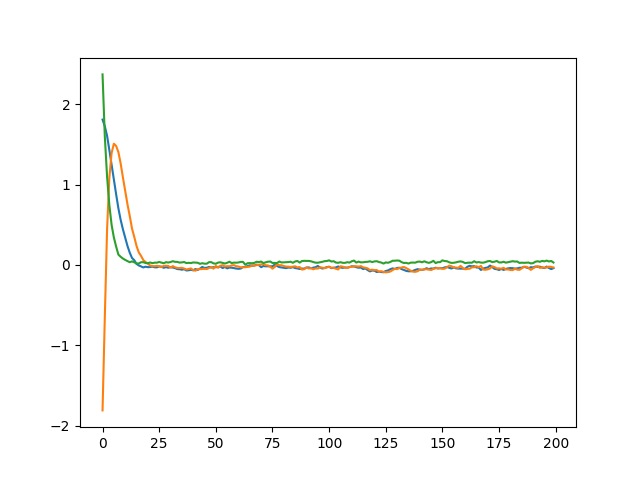}\\
                      {\small $|\lambda|=1, \arg\lambda=0$} & {\small $|\lambda|=1, \arg\lambda=0.1$} &
                      {\small $|\lambda|=1, \arg\lambda=0.2$} & 
                      {\small $|\lambda|=0.96, \arg\lambda=0.2$} & {\small $|\lambda|=0.92, \arg\lambda=0.2$} \\
  \end{tabular}}
  \caption{Examples of controlled dynamics by the proposed method with different target eigenvalues. The horizontal axis is time, and the vertical axis is values for each element of the measurement vectors. The two bottom rows show the absolute values and argument of the target eigenvalues.}
  \label{fig:controlled}
\end{figure*}

For the evaluation measurement, we used the mean absolute error between target eigenvalues
and eigenvalues of the Koopman operator of the controlled dynamics estimated by Hankel DMD in Eq.~(\ref{eq:r}).
Table~\ref{tab:error} shows the mean absolute errors averaged over ten experiments for each target eigenvalue,
where 50 simulation runs were used for evaluation for each experiment.
The target eigenvalue was selected from one of the following six values
$\bm{\lambda}\in\{1,e^{i0.1},e^{i0.2},e^{i0.3},0.96e^{i0.2},0.92e^{i0.2}\}$,
where there was one target eigenvalue for each experiment.
The proposed method achieved the best performance in most of the settings.
Figure~\ref{fig:controlled} shows examples of the controlled dynamics with the proposed method.
The proposed method adaptively controlled the dynamics depending on the target eigenvalue with different frequencies
and decay rates.
SL failed to control because it assumed linear dynamics although the given systems were nonlinear.
SN assumed nonlinear dynamics with the Koopman invariant subspace. However,
the system identification was performed without considering control,
and learned Koopman embeddings and dynamics were not optimal for control.
On the other hand, the proposed method learns the Koopman embeddings and dynamics
by maximizing the control performance.
The error by RL was small when the argument of the target eigenvalue was zero, i.e., the desired dynamics is not periodic.
However, it was large on the periodic desired dynamics.
It would be difficult for RL to control to be periodic since RL models the policy with a neural network without structure.
In contrast, the proposed method incorporates the pole placement method developed for controlling periodicity in the neural network 
based on the Koopman operator theory.

\section{Conclusion}

We proposed an end-to-end learning method for controlling the frequency and convergence rate of nonlinear dynamical systems
based on the Koopman operator theory and deep reinforcement learning.
With the proposed method,
dynamics in the Koopman invariant subspace and neural networks for embedding
are trained such that the control performance is improved.
We experimentally confirmed that the proposed method
achieved better performance than
model-free reinforcement learning, and 
two-step approaches of separated system identification and controller optimization.
Although our results are encouraging for controlling nonlinear dynamical systems,
several directions remain in which we must extend our approach.
First, we will clarify the conditions for the proposed method to control dynamical systems.
Second, we plan to improve our method by incorporating advanced reinforcement learning techniques.
Third, we want to develop control methods for nonlinear systems
by extending control methods for linear systems other than pole placement
using our framework.


\bibliographystyle{abbrv}
\bibliography{arxiv_dmd_acker}

\end{document}